\documentstyle[twoside,fleqn,espcrc2,epsf]{article}
\renewcommand{\floatsep}{5mm plus 2pt minus 0pt}
\textfloatsep=\floatsep
\intextsep=\floatsep
\newcommand{\Journal}[4]{{#1} {\bf #2}, #3 (#4)}

\newcommand{\NPB}{Nucl.~Phys.~B}
\newcommand{\PLB}{Phys.~Lett.~B}
\newcommand{\PRL}{Phys.\ Rev.~Lett.\ }
\newcommand{\PRD}{Phys.\ Rev.~D}
\newcommand{\ZPC}{Z.~Phys.~C}
\newcommand{\PRC}{Phys. Rep.~C}

\title{\bf Bag Picture of the Excited QCD Vacuum with Static 
           ${\mathbf Q}\bar{\mathbf Q}$ Source\thanks{Talk 
           presented by J.~Kuti. \vskip 1mm 
           \noindent This work was 
           supported by the U.S.~DOE, Grant No.\ DE-FG03-90ER40546}}
\author{ K.J.~Juge, J.~Kuti, and C.~J.~Morningstar
\address{Department of Physics,
        University of California at San Diego,
        La Jolla, California 92093-0319}
        }

\begin{document}

\begin{abstract}
The gluon excitations of the QCD vacuum are investigated in the presence
of a static quark-antiquark source. 
It is shown that the ground state
potential and the excitation spectrum of dynamical gluon degrees
of freedom, as determined in our
lattice simulations, agree remarkably well with model predictions
based on the diaelectric properties of the confining vacuum 
described as a dual superconductor. 
The strong chromoelectric field of the static $\mathrm Q\bar Q$
source creates a bubble (bag) in the condensed phase 
where weakly interacting gluon modes can be excited. Some features
and predictions of the bag model are presented and the chromoelectric
vortex limit at large quark-antiquark separation (string formation) is 
briefly discussed.
\end{abstract}

\maketitle

\section{Introduction}
The project which we initiated eighteen months ago to study
the gluon excitations of the QCD vacuum in the presence
of a static quark-antiquark source has two distinct goals. 
Our first objective
was to determine the spectrum of hybrid $\mathrm c\bar c g$
and $\mathrm b\bar b g$ states with
results reported in Ref.~\cite{JKM1}. 
Early predictions of these states in the
Born-Oppenheimer approximation were based on the bag picture 
of the QCD vacuum in Ref.~\cite{HHKR}, and additional
phenomenological observations were made
in Ref.~\cite{Ono}.

The investigation of gluon
excitations around a static quark-antiquark pair in the QCD vacuum
has important implications on our conceptual
understanding of the quark confinement mechanism about 
which very little is known at the present.
In this talk we will address this second goal within the context
of a simplified picture of the QCD vacuum which in the first tests
appears to agree surprisingly well with simulation results.

\section{Diaelectric vacuum and bag formation}

There is little doubt that some sort of a bag is formed when a 
static $\mathrm Q\bar Q$ pair is inserted in the physical vacuum
at a separation $\mathrm r\ll 1~fm$ where asymptotic freedom holds. 
The strong chromoelectric dipole field, 
$\mathrm E_\theta = \frac{2cos\theta}{\sqrt 3}g(r)/R^3$,
at a distance $\mathrm R$ from the dipole source,
suppresses the microscopic nonperturbative condensate before 
the field strength drops to some typical confinement
scale $\mathrm E_{critical} \sim \Lambda^2_{QCD}$
at a distance $\mathrm R_b$ which we will identify 
qualitatively as the bag radius of confinement (g(r) is the coupling
constant, or color charge). 
At the confinement scale, the perturbative vacuum bubble 
which is sustained by the strong dipole field
has to be replaced by the nonperturbative condensate of the physical vacuum.
Within the bubble (bag) we should be able to apply perturbation theory 
for gluonic corrections to the dipole field to recover the running Coulomb law.
The size of the bubble $\mathrm R_b$ can be
estimated from the relation
$\mathrm \frac{2}{\sqrt 3}g(r)/R_b^3 \sim \Lambda^2_{QCD}$.

In the bag model, as explained in Refs.~\cite{HK} and~\cite{HHKR}, 
we assume a simple confinement picture for the interface between the two
phases of the vacuum.
Inside the bag the chromoelectric vacuum permeability $\epsilon$ 
is set to one (perturbative vacuum). In the confining gluon
condensate of the physical vacuum $\epsilon=0$ 
is assumed (diaelectric vacuum) which is expected to
emerge from the microscopic theory of a dual nonabelian
superconductor in QCD.
A sharp boundary is assumed to separate the bag from the physical vacuum
with surface energy $\sigma$ per unit area 
and volume energy B per unit volume.
The value of B is related to the gluon vacuum condensate by the
relation $\mathrm B = -\frac{9}{32}\langle|\frac{\alpha_s}{\pi}F^a_{\mu\nu}
F^a_{\mu\nu}|\rangle$~\cite{Shifman}. Based on QCD sum rules and heavy $\mathrm Q\bar Q$
spectroscopy, the value of B is determined to be in the range 
$\mathrm B^{1/4} \sim 250-350~MeV$~\cite{Shifman}. 
The interface energy at the deconfining
transition temperature of quenched QCD has been determined in the 
$\mathrm \sigma\sim 5-10~MeV/fm^2$ range~\cite{HPRS} 
which is small in comparison
with the volume energy and neglected in the calculations we present.

\section{Adiabatic bag picture}
\begin{figure}
\begin{center}
\leavevmode
\epsfxsize=2.5in\epsfbox[0 0 534 568]{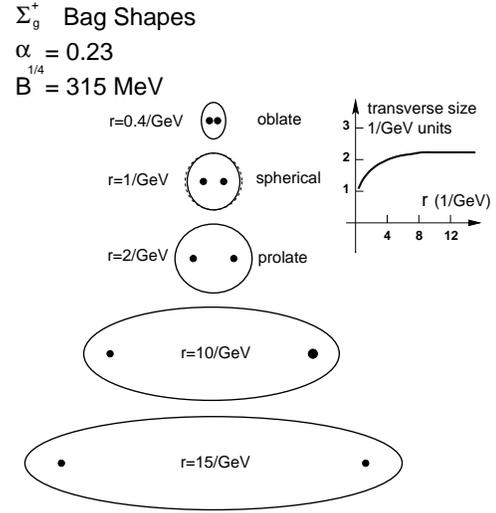}
\end{center}
\vskip -7mm
\caption{{\small The shape of the bag in the $\Sigma^+_g$
ground state is depicted 
as a function of quark-antiquark separation in the ellipsoidal
approximation. The black dots designate the locations of the Q and 
$\mathrm\bar Q$ sources. The oblate shape at small separation is
determined by the dominant dipole field.
The transverse size of the asymptotic vortex
solution is reached at $\mathrm r \sim 1~fm$ separation. }}
\label{fig:x1}
\end{figure}

\begin{figure}
\begin{center}
\leavevmode
\epsfxsize=1.9in\epsfbox[0 0 380 443]{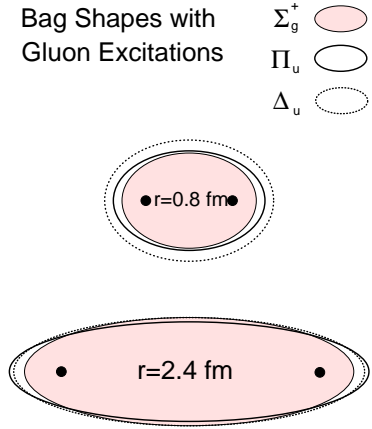}
\end{center}
\vskip -7mm
\caption{{\small Bag shapes with gluon excitations are compared
with the shape of the bag in its ground state at two different
quark-antiquark separations.}}
\label{fig:x3}
\end{figure}

The adiabatic method we apply here is a variational principle
for the total energy of the bag (for details we refer to
Refs.~\cite{HHKR} and ~\cite{HK}). In the results we present here
an ellipsoidal shape is used in the variational calculations which
is adequate within a few percent accuracy.
With an effective coupling constant $\alpha_s$ inside the bag, and 
with the choice of B given in Fig.~\ref{fig:x1}, we first solve
the bag equations in Coulomb gauge for the ground state when 
dynamical gluons are
not excited. The variation of the minimal bag shape as a function
of the $\mathrm Q\bar Q$ separation is depicted in 
Fig.~\ref{fig:x1}. 
\epsfverbosetrue
Various bag shapes in the presence of gluon excitations are shown 
in Fig.~\ref{fig:x3}. The notation for the gluon quantum numbers is explained
in Ref.~\cite{JKM1}.

\begin{figure}
\begin{center}
\leavevmode
\epsfxsize=2.9in\epsfbox[18 244 592 718]{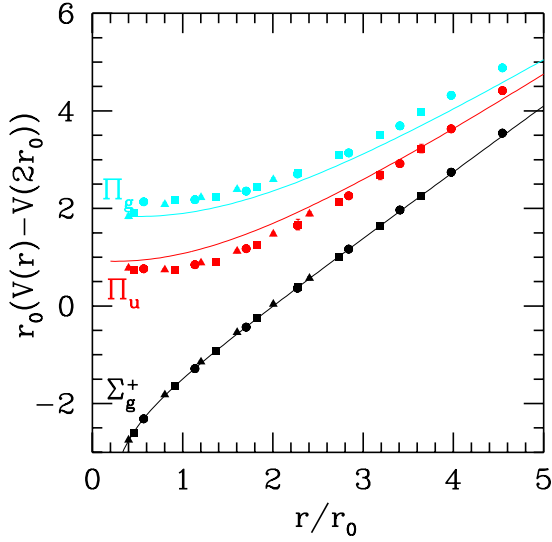}
\end{center}
\caption{{\small The bag model
predictions for the CP even $\mathrm V_{\Pi_g}(r)$
and CP odd $\mathrm V_{\Pi_u}(r)$ excitations with 
$\mathrm\Lambda = 1$ angular
momentum projection on the $\mathrm Q\bar Q$ axis
are depicted as the solid curves in units of the hadronic scale
parameter $\mathrm r_0$ (defined in \cite{JKM1})
against the quark-antiquark separation $\mathrm r$.
}}
\label{fig:bag1}
\end{figure}

\begin{figure}
\begin{center}
\leavevmode
\epsfxsize=2.9in\epsfbox[18 244 592 718]{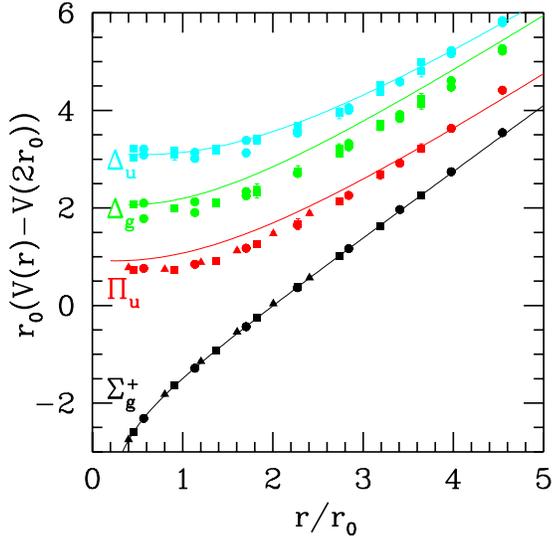}
\end{center} 
\caption{{\small The CP even $\mathrm V_{\Delta_g}(r)$ 
and CP odd $\mathrm V_{\Delta_u}(r)$ $\mathrm\Lambda = 2$ excitations
are depicted together with $V_{\Sigma^+_g}(r)$ 
and $V_{\Pi_u}(r)$.}}
\label{fig:bag2}
\end{figure}

The bag model in the adiabatic approximation predicts the gluon excitations
without free parameters. Fig.~\ref{fig:bag1} and Fig.~\ref{fig:bag2}
compare the bag model predictions with our simulation results. The 
agreements are quite remarkable.
\epsfverbosetrue

\epsfverbosetrue

\epsfverbosetrue

\section{Chromoelectric vortex (string) limit}

\begin{figure}
\begin{center}
\leavevmode
\epsfxsize=1.9in\epsfbox[0 0 477 640]{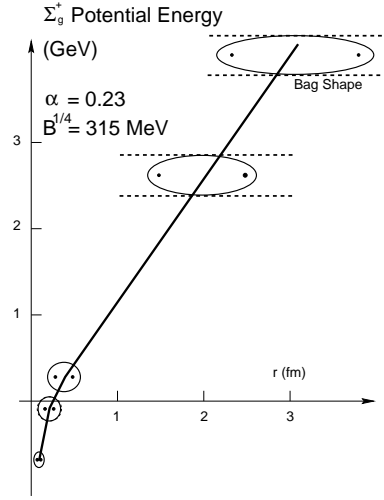}
\end{center}
\vskip -10mm
\caption{{\small The shape of the bag in the $\Sigma^+_g$
ground state is depicted 
as a function of quark-antiquark separation.
The exact vortex solution is shown by the dashed lines.
}}
\label{fig:x2}
\end{figure}

In the adiabatic approximation there is an exact vortex solution
to the bag equations which describes a homogeneous chromoelectric
flux with an intrinsic radius 
$\mathrm R_{vortex}=(8\alpha_s/3\pi B)^{1/4}$, where $\alpha_s$ and
B are the only two parameters of the model. The vortex energy per unit length
(string tension, or slope of the linear part of the
$\mathrm Q\bar Q$ potential) is given by 
$\mathrm \kappa=\sqrt{32\pi\alpha_s B/3}$. The vortex limit
is illustrated in Fig.~\ref{fig:x2}.
As it was shown in Ref.~\cite{HK} this vortex has massive intrinsic
gluon excitations along the ``waveguide" of the vortex
and collective string excitations which correspond to the low energy
Goldstone modes of the soliton~\cite{Luscher}.

The bag model provides a first attempt for a unified
low energy effective theory
to capture the physics of quark confinement at small {\em and} large
$\mathrm Q\bar Q$ separations. Effective string models focus on the long range
part of the picture~\cite{PS}. It remains a challenge to understand
the more detailed connection between the two approaches.
\epsfverbosetrue

\end{document}